# Remote epitaxy beyond polarity

Ching-Tai Fu[1†], Pei-Jan Hung[1,4†], Xudong Li[3†], Xiaolong Zhu[1†], Yu Han[1†], Sayantan Mahapatra[5], Zhiheng Zhao[1], Qingsong Fan[1], Xubing Wu[1], Qizhang Li[1], Chenxi Sui[1], Zirui Zhou[2], Ting-Hsuan Chen[1], Cheng-Hao Lei[1], Ivan Kuzmenko[4], Xiaobing Zuo[4], Byeongdu Lee[4], Alexander S. Filatov[2], Jeffrey R. Guest[5], Fengyuan Shi[6], Yuzi Liu[5], Hua Zhou[4]*, Yunfeng Shi[3]*, Po-Chun Hsu[1]*

Affiliations:

[1]Pritzker School of Molecular Engineering, University of Chicago, Chicago, IL, USA

[2]Department of Chemistry, James Franck Institute, University of Chicago, Chicago, IL, USA

[3]Department of Materials Science and Engineering, Rensselaer Polytechnic Institute, Troy, NY, USA

[4]X-ray Science Division, Advanced Photon Source, Argonne National Laboratory, Lemont, IL, USA

[5]Center for Nanoscale Materials, Argonne National Laboratory, Lemont, IL, USA

[6]Electron Microscopy Core, University of Illinois Chicago, Chicago, IL, USA

†These authors contributed equally

*Correspondence: Po-Chun Hsu (pochunhsu@uchicago.edu), Yunfeng Shi (shiy2@rpi.edu),
Hua Zhou (hzhou@anl.gov)

## Abstract

Remote epitaxy through a monolayer two-dimensional material-covered substrate establishes a crystallographic registry across the van der Waals (vdW) surface that enables the epitaxial growth, lift-off and transfer of single-crystalline films[1,2]. A central belief in remote epitaxy is that the substrate facilitating the phenomenon must be a material with strong ionicity, as the interatomic electrostatic potential fluctuation in covalent and metallic materials is substantially attenuated by two-dimensional materials[3]. Here, we show remote epitaxy is possible when the substrate is a metallic or covalently bonded material and experimentally demonstrate non-polar remote homo- and heteroepitaxy across a wide range of material systems, including both metals and semiconductors. The achieved non-polar remote interactions are designed and engineered by harnessing substrate conductivity and vicinal surface step-edge density. These findings indicate that remote epitaxy is universal and applicable to ionic, metallic, and covalent materials, expanding its capabilities and stimulating a plethora of new fundamental scientific questions about the mechanism of remote epitaxy.

## Main

Remote epitaxy–the epitaxial growth and release of epilayers on a semitransparent, two-dimensional material-covered substrate–offers the production of single-crystalline, freestanding films for seamless integration onto arbitrary platforms[1–10]. Successful demonstrations of remote epitaxy leverage strong electrostatic potential fluctuation of substrates with bonding polarity, through which the crystallographic registry can be transmitted across monolayer and multilayer two-dimensional interlayers, thereby enabling epitaxial growth. The weak van der Waals interaction between the epilayer and substrate allows facile lift-off and transfer to target substrates. A rich variety of remote epitaxial growth–for example, III-V semiconductors[11–16], metal oxides[17–22], complex oxides[5,23,24], and metal-halide perovskites[25,26]–have been realized, with remote epilayers demonstrating device applications and functionalities including light-emitting diodes, optical cavities, flexible photodetectors, infrared bolometers, and electronic skin.

In remote epitaxy, theoretical and empirical observations have identified the pivotal role of bonding polarity in governing remote interactions that guide adatom nucleation and growth through field potentials penetrating up to a few layers of graphene. This mechanism enables highly ionic materials, such as LiF, when used as a substrate, to achieve remote epitaxy up to three layers of graphene[3]. When the substrate features non-polar bonding, such as Si and Ge, remote epitaxy is no longer observed even through a single layer of graphene. In combination with polarized bonds, researchers have examined defects such as dislocations in polar GaN substrates as additional mediators to demonstrate remote epitaxy through 2-7 nm of amorphous carbon[27]. Beyond intrinsic polarity, material defects, which can be universally designed and engineered to enhance epilayer–substrate interaction, could also potentially be used to expand the remote epitaxy landscape.

Since the emergence of remote epitaxy, extending the concept to material systems with non-polar bonding has been highly sought after by the epitaxy community. These materials of weak polarity are generally covalently or metallically bonded, such that their electrostatic potential profiles are substantially attenuated across two-dimensional interlayers[3,28,29]. In principle, remote epitaxy through substrates with non-polar bonding, given the weak ionic character of the underlying material, is extremely difficult to realize. Over many remote epitaxy attempts using substrates with non-polar bonding, which include growth on Si and Ge substrates through monolayer graphene (ML-graphene) by metal–organic chemical vapor deposition[30,31] and the electrodeposition of Zn on graphene-Cu foil[32,33], films result in polycrystalline domains and micrometer-size segmented islands with discontinuities across the surface. Consequently, a viable strategy for realizing remote epitaxy of material systems with non-polar bonding has yet to be established.

Here, we demonstrate a universal approach for achieving non-polar remote epitaxy by harnessing the conductivity and step-edge density of metallic and covalently bonded substrates under ambient liquid-phase epitaxy. Conductive substrates with non-polar bonding cut at high vicinal angles contain densely distributed step-edges, which we have found effective in establishing epitaxial registry between the epilayer and the underlying substrate through ML-graphene. Non-polar remote epitaxy is prominently observed across multiple material systems, including Au film, $Cu_2O$ film, and ZnO nanorods on vicinal Au substrates, Cu film on Si wafer, and Cu film on Cu foil, suggesting the universality of this method. These epilayers are large, single-crystalline, and can be released as freestanding foils by facile lift-off from the vdW interface. Density functional theory modeling indicated that conductive step-edges directly underneath ML-graphene induce localized electron depletion that results in the non-polar electrostatic interactions required for remote epitaxy. These results identify substrate conductivity and step-edge density as critical key enablers for extending remote epitaxy beyond the conventional polarity-driven framework.

## Realizing non-polar remote epitaxy

Figure 1a shows the conventional remote epitaxy, using GaAs through ML-graphene on a polar GaAs substrate as an example. GaAs possesses substantial ionic character, and its surface potential can easily penetrate through ML-graphene. The resulting electrostatic potential fluctuations at the graphene surface guide GaAs adatoms to follow the atomic registry of the underlying substrate for epitaxial growth. By contrast, on substrates with non-polar bonding, the nuclei–substrate interaction is barely preserved (Fig. 1b). When a weakly ionic Ge substrate is coated with ML-graphene, the substrate potential is effectively screened. In the absence of sufficient electrostatic guidance, nuclei form with random orientations, resulting in polycrystalline Ge. To address this limitation, as shown in Fig. 1c, we present our strategy to bypass the intrinsic constraints of substrate materials and engineer the electrostatic potential in non-polar systems to achieve remote epitaxy. We first prepare vicinal Au substrates by growing 20-nm-thick epitaxial Au on Si wafers off-cut by 4° from the [111] axis, followed by ML-graphene coating. 20-nm-thick Au epilayer is then grown on graphene-covered vicinal Au by electrochemical liquid-phase epitaxy at room temperature. Vicinal substrates contain a series of atomically high steps, which we expected would create specific step-edge nucleation sites beneath ML-graphene. Compared with ML-graphene, Au, which has much higher electron affinity, induces localized electron depletion in graphene that generates abrupt potential fluctuations at the graphene–step-edge sites. The resulting potentials penetrate through ML-graphene and emerge across its surface, enabling non-polar remote epitaxy.

Single-crystal Au epilayers are obtained from this method. Scanning electron microscopy (SEM) images of the Au/ML-graphene/vicinal-Au sample shown in Fig.1d indicated smooth and uniform morphology and enabled the corresponding electron backscatter diffraction (EBSD) inverse pole figure mapping that showed a (111) crystallographic orientation in the out-of-plane direction (Fig. 1e). The X-ray diffraction (XRD) pattern of Au/ML-graphene/vicinal-Au (Fig. 1f) showed a series of (111) peaks under the logarithmic scale, and the out-of-plane epitaxial relationship can be described as Au (111) || Au (111). The in-plane epitaxial relations were confirmed to be Au $[10\bar{1}]$ || Au $[10\bar{1}]$ by Au (220) X-ray pole figure (Fig. 1g) which featured the characteristic six-fold symmetry of the twinned (111) crystalline domains rotated 60° in-plane to each other.

To directly verify the remote epitaxial relationship between the Au epilayer and underlying Au substrate, we performed cross-sectional STEM to atomically resolve the Au/ML-graphene/vicinal-Au interface. High-angle annular bright-field scanning TEM (STEM) (Fig. 1h and 1j) and annular dark-field STEM (Fig. 1i) imaging showed that the Au epilayer is remote-epitaxially grown with the Au substrate across a continuous ML- graphene. The ML-graphene is clearly visible and uniform across the Au epilayer–Au substrate interface, and the measured vdW gap between the Au epilayer and the Au substrate is approximately 4.6 Å. The corresponding electron energy-loss spectroscopy (EELS) (Fig. 1k) mapping of the Au/ML-graphene/vicinal-Au interface further confirmed the presence of graphene through the acquired carbon K-edge. STEM analysis confirms the successful realization of remote epitaxy through ML-graphene in non-polar systems.

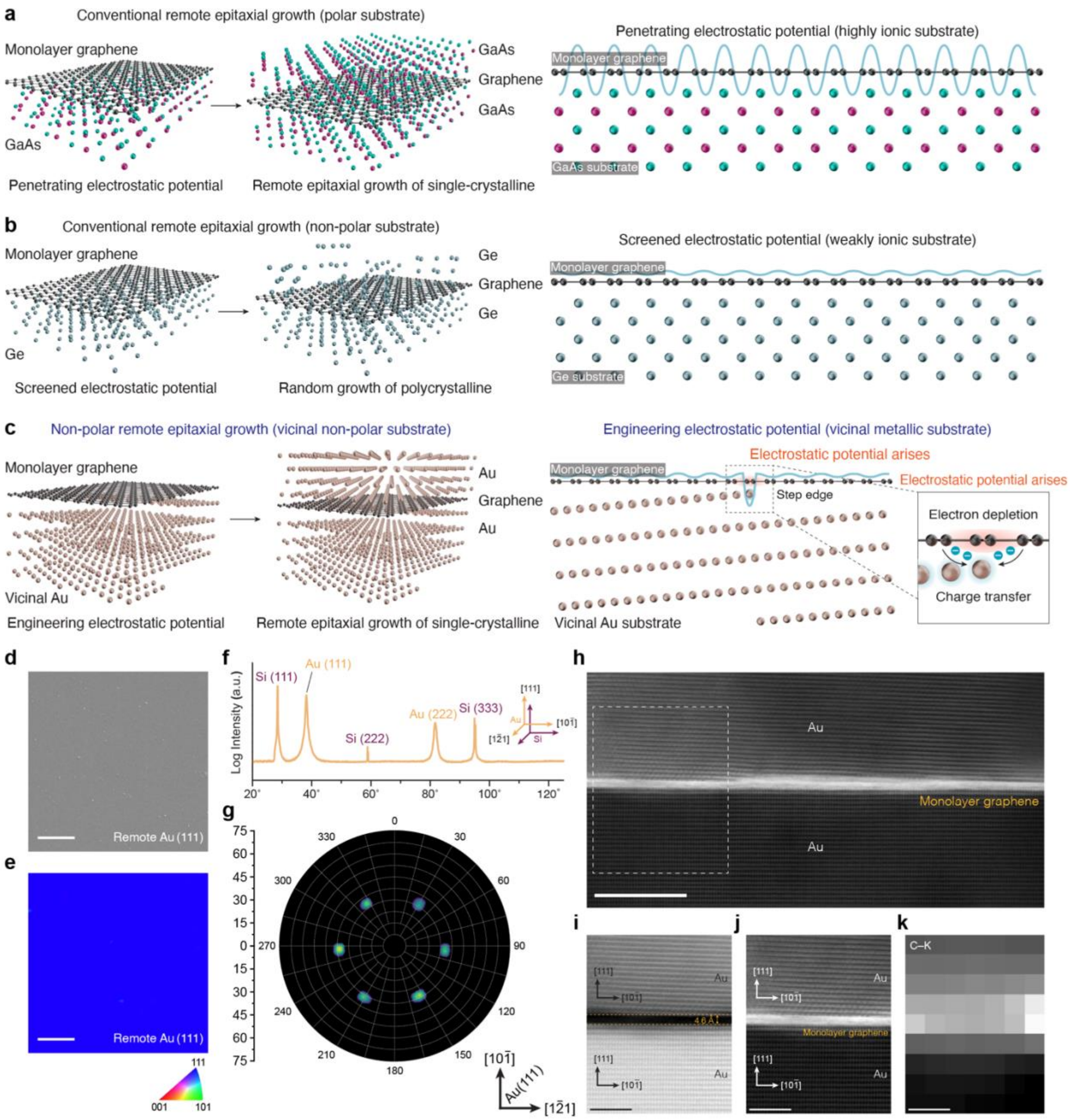

**Fig. 1. Remote epitaxy beyond polarity and the remote epitaxial growth of Au film on ML-graphene/vicinal-Au. a**, Schematic of the conventional remote epitaxy process using a polar GaAs substrate. **b**, Schematic of an unsuccessful non-polar remote epitaxy (polycrystalline growth) using a graphene-coated, non-polar Ge substrate. **c**, Schematic of the vicinal substrate strategy to address the limitations of conventional remote epitaxial growth to achieve non-polar remote epitaxy. **d**,**e**, SEM image (**d**) and the corresponding EBSD inverse pole figure of Au/ML-graphene/vicinal-Au (**e**). **f**, XRD pattern (log-scale intensity) for Au film on ML-graphene/vicinal-Au. **g**, X-ray pole figure from the Au (220) facets. **h**, Annular bright-field STEM image showing the Au/ML-graphene/vicinal-Au interface. **i**-**k**, Zoomed-in high-angle annular dark-field (HAADF)

image (**i**), annular bright-field STEM image (**j**), and electron energy-loss spectroscopy (EELS) of image **h** (**k**). Scale bars, 5 μm (**d**-**e**), 5 nm (**h**), 2 nm (**i**-**k**).

### Atomic-scale modeling of graphene–step-edge interaction in non-polar remote epitaxy

To investigate whether conductive vicinal steps beneath ML-graphene can generate the electrostatic potential required for non-polar remote epitaxy, we performed density functional theory (DFT) calculations (Methods). We expected that substrate conductivity and varying the vicinal angles of the substrate (and thereby tuning the terrace width and step-edge density) would systematically modulate the metal ion–substrate interaction in material systems with non-polar bonding. Subsequently, we evaluated the electrostatic potential energy depths (in terms of one electron, consistent with VASP convention) formed across ML-graphene-covered vicinal substrates, including Au and Cu as conductive metallic substrates, GaAs as a substrate with polar bonding, and Si as a substrate with non-polar bonding (Supplementary Note 1). Our analyses show that, in the presence of step-edges, ML-graphene-covered metallic substrates exhibit twice the electrostatic potential energy depth, while the electrostatic potential energy depths from GaAs and Si are substantially attenuated across the graphene interlayer (Fig 2a and Supplementary Fig. 1, summarized in Table 1). We further calculated the corresponding flat substrate configurations, including ML-graphene-covered Au, Cu, GaAs and Si (Supplementary Note 2). As shown in Supplementary Fig. 2, in the absence of step-edges, the electrostatic potential energy fluctuations of substrates with non-polar bonding are effectively screened by ML-graphene, whereas those of substrates with polar bonding remain preserved, consistent with previous theoretical observations of remote epitaxy[3]. These results indicate that conductive step-edges are required to generate the electrostatic potential needed for non-polar remote epitaxy (Table 1). In addition to the electrostatic potential energy profile characterizing metal ion-substrate interaction, we investigated the potential energy profiles governing the remote interactions of neutral Au and Cu adatoms with their corresponding vicinal metallic substrates (Au and Cu, respectively) across ML-graphene (Supplementary Note 3). Our calculations identify graphene sites located directly above the underlying step-edges as the most energetically favorable post-reduction adsorption sites. In the lowest-energy configuration, the neutral atom resides on ML-graphene directly above a buried step-edge, indicating that conductive step-edge-induced potential energy landscape guides adatom positioning and promotes atomic registry with the underlying substrate (Supplementary Fig. 3).

More importantly, to further identify the fundamental origin of conductive step-edge-induced electrostatic potential energy landscape, we examined charge redistribution at the graphene-vicinal substrate interface (Supplementary Note 4). Our analysis reveals that charge transfer occurs directly at the graphene–conductive step-edge interface. Figure 2a shows that Au and Cu step-edges actively withdraw charge from graphene (with charge depletion and accumulation represented in blue and red, respectively, in the figure). In contrast, charge transfer at the graphene–step-edge interface is negligible for GaAs or Si (Fig. 2b). Together, these theoretical insights demonstrate that conductive step-edges induce localized electron depletion in ML-graphene, which gives rise to periodic electrostatic potential energy profile near the step-edges and thereby facilitates non-polar remote epitaxy.

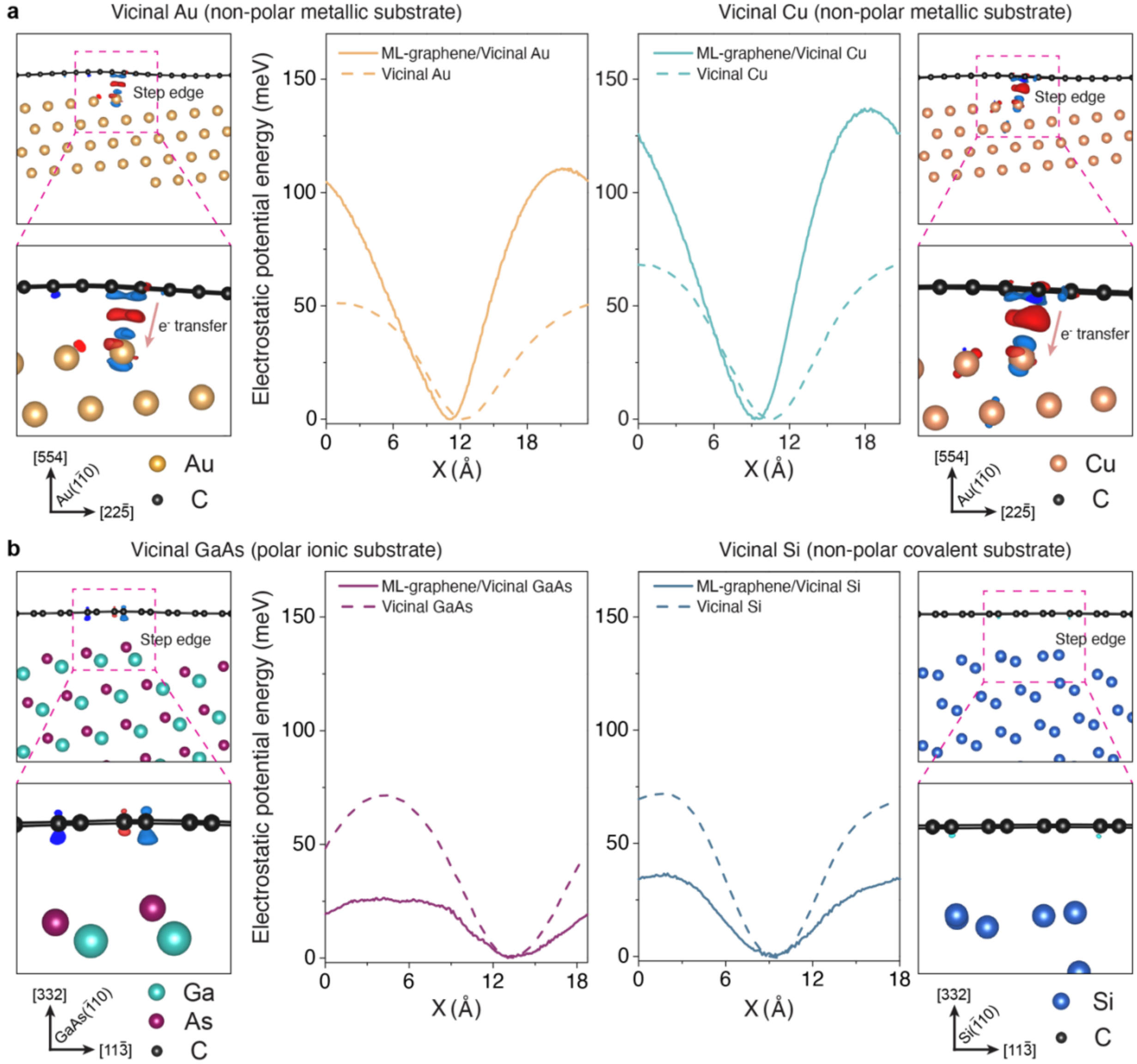

**Fig. 2. Electrostatic potential energy profile of step-edge-induced atomic interactions in non-polar remote epitaxy. a**,**b**, Visualized charge transfer redistribution and electrostatic potential energy profiles analyzed at the graphene–step-edge interface of (**a**) ML-graphene/vicinal Au (left) and ML-graphene/vicinal Cu (right) and (**b**) ML-graphene/vicinal GaAs (left) and ML-graphene/vicinal Si (right). In the three-dimensional electron-transfer isosurfaces, red and blue denote charge accumulation (electron inflow) and charge depletion (electron outflow), respectively. In metallic systems, the blue areas directly beneath ML-graphene at the step-edge interface represent charge-depleting regions (belonging to the π-bond of graphene), whereas the red areas indicate the enlarged charge-accumulating region (belonging to the metallic bond of Au or Cu). The red arrows indicate the direction of charge transfer. The charge accumulation and depletion regions are substantially weaker at the graphene–step-edge interface of GaAs and Si.

| System | | Electrostatic potential energy depth (meV) | | | |
|---|---|---|---|---|---|
| **Graphene overlayer** | **Substrate step edge** | **Au** | **Cu** | **GaAs** | **Si** |
| ✓ | ✓ | **110.5** | **138.1** | **25.9** | **35.2** |
| — | ✓ | **51.4** | **68.6** | **71.3** | **72.8** |
| ✓ | — | **12.7** | **12.9** | **19.8** | **9.9** |
| — | — | **2.6** | **2.3** | **19.4** | **1.4** |

**Table 1.** Demonstration of non-polar remote interaction from conductive vicinal surfaces

**Epitaxial quality of Au grown on monolayer graphene**

To further validate the remote epitaxial quality of Au epilayer grown on ML-graphene, we conducted systematic synchrotron-based X-ray scattering characterization of the remote Au epilayer. Three-dimensional reciprocal space mapping (3D RSM) captures the full distribution of diffracted intensity around a reciprocal lattice point, providing a sensitive measure of epitaxial quality. 3D RSM of the specular, Au (111) peak (Fig. 3a and Supplementary Note 5) after non-polar remote epitaxy confirms out-of-plane coherence and epitaxial relationship between the Au epilayer and Au substrate, with the corresponding slices along the H-L, K-L and H-K reciprocal planes (Fig. 3b-d) featuring clear Laue fringes.

To determine the epitaxial quality of Au epilayer grown by non-polar remote epitaxy, we further examined crystal mosaicity along three independent crystallographic directions–specular, off-specular, and orthogonal. First, we focused on the specular [111] direction (out-of-plane), which probed both Au epilayer and Au substrate. Specular rocking curve (Fig. 3e) showed a narrow peak with a full width at half maximum (FWHM) of 1.170°, indicating exceptional alignment in the out-of-plane direction between the remote epilayer and substrate. To elucidate the crystalline characteristics only in the Au epilayer, we then performed rocking curves at grazing incidence (Supplementary Note 6 and Table S2.) for the off-specular (in-plane) and orthogonal (perpendicularly in-plane) directions. By varying the incident angle, we adjusted the X-ray penetration and scattering depths and analyzed crystal mosaicity at defined depths from the surface of the remote Au epilayer. Off-specular rocking curve (Fig. 3f) in the $[1\bar{1}1]$ direction showed consistent FWHM, averaging 2.554˚± 0.023˚ across all grazing angles (corresponding to 6–81 nm from the epilayer's top surface), which revealed strong in-plane crystallographic correlation between the substrate and epilayer through ML-graphene. In the $[2\bar{2}0]$ direction, orthogonal rocking curve (Fig. 3g) showed that the FWHM, averaging 2.477˚± 0.045˚, was again consistent across all grazing angles (6–87 nm from the epilayer's top surface), verifying that the epitaxial in-plane registry of Au single-crystal grown on the ML-graphene/vicinal-Au is well-maintained. Taken together, these observations confirm a coherently oriented single-crystal lattice extending from the substrate through ML-graphene and into the epilayer, providing direct evidence of non-polar remote epitaxy in obtaining high-quality single-crystals.

To validate our conductive step-edge strategy, we resolved the single-crystallinity of Au grown on ML-graphene using 3D RSM and correlated with the step-edge densities using scanning tunneling microscopy (STM) on both vicinal and nominally flat substrates. Single-crystal Au substrates grown on Si wafers with varying vicinal angles define the conductive, non-polar surface beneath ML-graphene; the varying miscut angles systematically tune the terrace width and step-edge density of the Au surface (Supplementary Fig. 6 and Supplementary Fig. 7). In the nominally flat Si substrate, which exhibited an experimentally measured off-cut of 0.20˚, two-dimensional projection of 3D RSM (Fig. 3h and Supplementary Note 5) showed $\{1\bar{1}1\}$ and {200} diffraction spots with the presence of Debye–Scherrer rings, indicating the coexistence of preferentially oriented crystallites and non-epitaxial polycrystalline domains. Scanning tunneling microscopy (STM) measurements (Fig. 3i) revealed terrace widths of ~24 nm on the 0.20˚ nominally flat substrate, corresponding to a relatively low step-edge density. In stark contrast, the two-dimensional projection of 3D RSM of Au films grown on 4˚ vicinal Si substrate (Fig. 3j), with a measured off-cut of 3.90˚, exhibited pronounced Bragg spots from the $\{1\bar{1}1\}$ and {200} reflections and a much higher single-crystal ratio (Supplementary Fig. 8, Supplementary Fig. 9, and Table S3), confirming the single-crystalline nature of Au films grown by non-polar remote epitaxy. STM measurements further showed terrace widths below 3 nm on the 4° vicinal substrate, revealing much narrower terraces and substantially higher step-edge density (Fig. 3k). These results demonstrate the systematic correlation between step-edge densities in substrate and the epilayer quality above ML-graphene, providing evidence of the pivotal role of dense vicinal steps in non-polar remote epitaxy. The complete set of $\{1\bar{1}1\}$ poles of single-crystal Au grown on dense, vicinal stepped Au are shown in the corresponding 3D RSM (Fig. 3l and Supplementary Video 1). Consistent with previous XRD results, the combination of a vicinal substrate with ML-graphene induces remote interactions between the substrate and epilayer, as reflected by both specular (out-of-plane) and off-specular (in-plane) poles in the full 3D RSM. These data confirm the single crystal growth of non-polar remote epitaxy and validate the epitaxial relationship as Au (111) || Au (111) in the out-of-plane direction and Au $[10\bar{1}]$ || Au $[10\bar{1}]$ in the in-plane direction.

During electrodeposition, an electrical double layer forms at the interface between graphene and the electrolyte, where cations and anions are tightly packed and strong electrical field exists. Because of the small intermolecular distance between the first (cationic) and the second (anionic) layers, the electrical field can be hundreds of MV/m even with only a few volts of operational voltage[34]. To elucidate the relationship, we deposited Au on ML-graphene/vicinal-Au at various applied potentials (-3.7 V, -2.8 V, -1.9 V, -1.0 V, and -0.1 V versus Ag/AgCl). Grazing-incidence wide-angle X-ray scattering (GIWAXS) patterns (Fig. 3m and Supplementary Note 6), collected at a small grazing angle (20 nm from the surface of the remote film), showed polycrystalline Au at both more negative potentials (-3.7 V and -2.8 V) and less negative potentials (-1.0 V and -0.1 V), as evidenced by Debye–Scherrer rings accompanied by faint diffraction spots. At more negative potentials (−3.7 V and −2.8 V), the polycrystallinity can be explained by the excessive growth rates due to strong electrochemical driving forces, kinetically preventing Au adatoms from relaxing into the proper epitaxial registry[35]. This phenomenon is consistent with classical thin film growth theory, which predicts the loss of epitaxial relationship when the adatom burial time is significantly shorter than the diffusion time[36,37].

On the other hand, the polycrystallinity at less negative potentials (−1.0 V and −0.1 V) is counterintuitive, as slow growth rate usually leads to higher epitaxy quality[38]. Instead, this result implies, during electrodeposition, the applied potential (and the consequent electrical double layer) is conducive to the epilayer-substrate interaction—a phenomenon that is particularly essential when the epilayer and the substrate are separated by the graphene. In striking contrast, the Au film electrodeposited at -1.9 V exhibited pronounced off-specular Au $(1\bar{1}\bar{1})$ and Au (200) Bragg spots, together with an emerging streak of crystal truncation rod connecting the two reflections, indicating a highly ordered, single-crystalline Au film with an atomically smooth surface (Fig. 3m). These features reveal that at -1.9 V, the electric double layer provides an optimal driving force that allows Au adatoms to adopt the remote epitaxial registry.

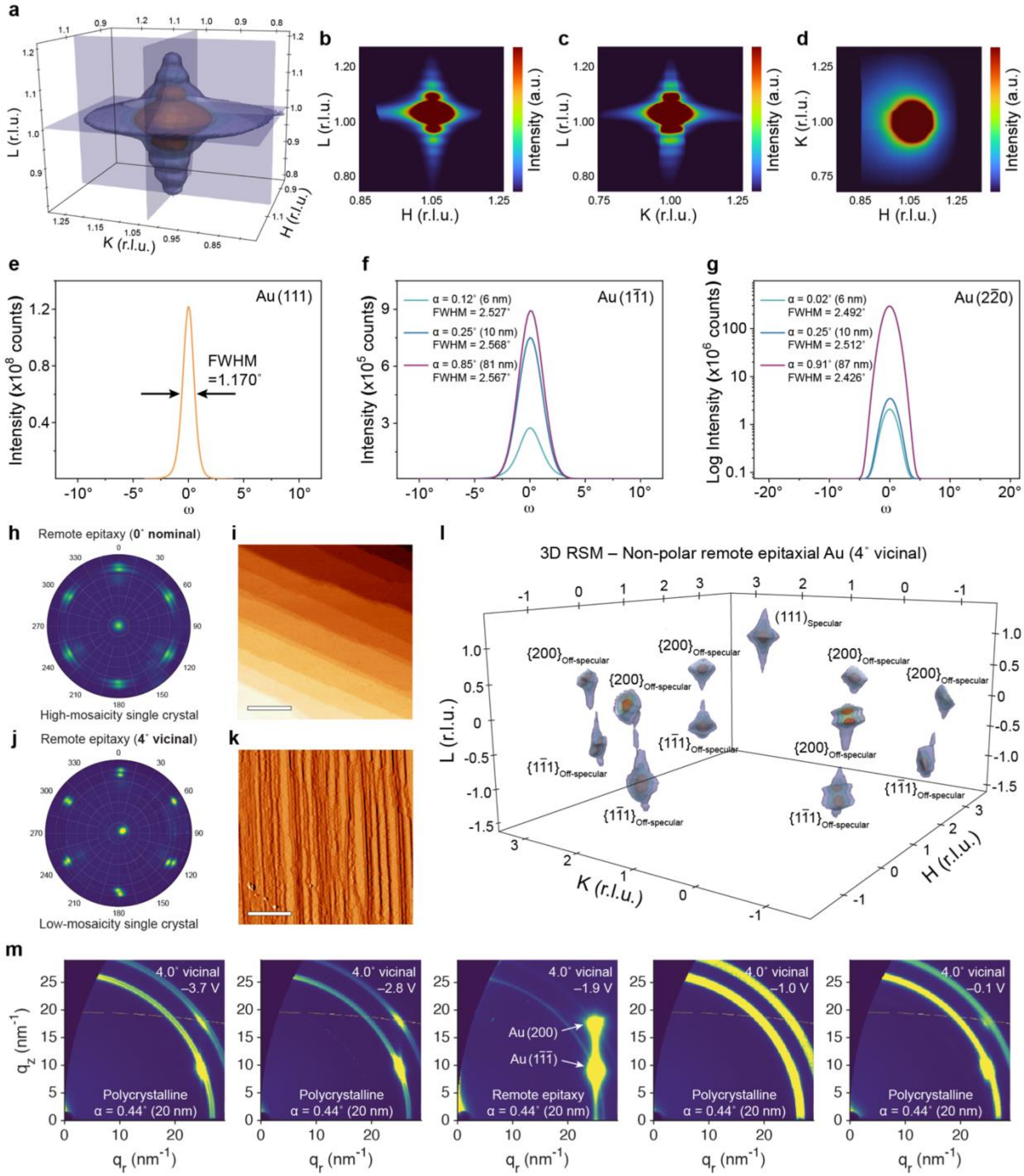

**Fig. 3. Synchrotron X-ray scattering characterizations of remote-epitaxial Au grown on ML-graphene/vicinal-Au. a-d**, Synchrotron X-ray 3D RSM of the Au specular (111) peak (**a**) with corresponding slices along the H-L (**b**), K-L (**c**), and H-K (**d**) reciprocal planes. **e,** Specular rocking curve of 20-nm thick Au epilayer grown on ML-graphene/vicinal-Au with a FWHM of 1.170°, confirming the high crystalline quality of the Au epilayers. **f**,**g**, Grazing-incidence rocking curves of off-specular Au ($1\bar{1}1$) peak with selected incident angle α = 0.12, 0.25, and 0.85° (**f**) and orthogonal Au ($2\bar{2}0$) peak with selected incident angle α = 0.02, 0.25, 0.91° (**g**). Both off-specular

and orthogonal rocking curves exhibit consistent FWHM across all incident angles, confirming coherent crystallinity enabled by non-polar remote epitaxy. **h**,**i**, Two-dimensional projection of the 3D RSM of Au epilayer grown on a nominally flat (vicinal angle only 0.2˚) Au substrate (**h**) and the corresponding STM topography of the underlying Si surface ($V_S$ = 1.92 V, $I_t$ = 343 pA) (**i**). **j**,**k**, Two-dimensional projection of the 3D RSM of Au epilayer grown on 4.0˚ vicinal Au substrate (**j**) and the corresponding derivative STM topography of the underlying Si surface ($V_S$ = 1.10 V, $I_t$ = 155 pA) (**k**). Au epilayer grown on the low-step-density 0.2° substrate exhibits high mosaicity and obvious polycrystalline domains, whereas the Au epilayer grown on the high-step-density 4.0° vicinal substrate shows low mosaicity and exceptional single-crystallinity. **l**, 3D RSM of Au grown on of 4˚ vicinal substrate, corresponding to the 2D projection in **j**. **m**, GIWAXS patterns of Au grown on ML-graphene/vicinal-Au(111) with controlled polarizations of -3.7, -2.8, -1.9, -1.0, and -0.1 V at an incident angle of 0.44˚. Scale bars, 40 nm (**i**,**k**).

**Universality of non-polar remote epitaxy on covalent and metallic bonded substrates**

With vicinal conducting substrates and the electrical double layer conducive to non-polar remote epitaxy, we further our study to other material systems to demonstrate its generality. With judiciously designed electrochemical parameters, we successfully demonstrate non-polar remote epitaxy of single-crystalline $Cu_2O$ (111) film on Au (111) (Fig. 4a-d), ZnO (002) nanorods on Au (111) (Fig. 4e-h), Cu (100) film on extrinsic Si (100) (by a 45° rotational coincidental lattice matching) (Fig. 4i-l), and Cu (100) film on Cu (100) foil (Fig. 4m-p). These films are single-crystalline in nature and exhibit epitaxial registry with their underlying substrate, as confirmed by EBSD inverse pole figure mappings (Fig 4a,i,m), XRD patterns (Fig. 4b,f,j,n) and X-ray pole figures (Fig. 4c,d,g,h,k,l,o). The remote epitaxy of various materials through metallic and covalent substrates shows the universality and strong correlation between conductive step-edges and non-polar remote interactions.

In addition to designing and engineering vicinal single-crystal wafers to facilitate non-polar remote epitaxy, commercially available graphene-Cu foil directly provides an ideal stepped, conductive platform for large-scale application of non-polar remote interactions. Graphene grown via chemical vapor deposition (CVD) on Cu foil is predominately monolayer and continuous across the surface[39,40]. During high-temperature annealing and graphene growth, the underlying graphene-covered Cu foil surface reconstructs to form pronounced step bunches, which serve as an effective template for non-polar remote epitaxy (Supplementary Fig. 10). Inverse pole figure mapping (Fig. 4m) and XRD pattern of Cu on ML-graphene/Cu-foil (Fig. 4n) showed exceptional orientational purity, detecting only a collection of Cu (100) peaks. The out-of-plane epitaxial relationship can be described as Cu (001) || Cu (001). The X-ray pole figure of Cu {111} planes (Fig. 4o) exhibited four-fold symmetry, verifying in-plane epitaxy relation was Cu [100] || Cu [100]. In addition, we demonstrate that the remote Cu epilayer can be released by mechanical exfoliation, demonstrating the feasibility of release and transfer by non-polar remote epitaxy (Fig. 4p and Supplementary Video 2). To our knowledge, freestanding single-crystal metal films have previously been fabricated only through energy-intensive post-processing[41,42] or using Si wafer-based growth platforms[43,44]. In contrast, our approach directly produces freestanding single-crystal

Cu foils, demonstrating both the broad substrate compatibility and practical efficacy of our method. Here, non-polar remote epitaxy offers promising advancements in single-crystal foil technology and significant potential for various industrial applications.

In conclusion, we have demonstrated a fundamentally intriguing phenomenon as well as a powerful strategy to achieve non-polar remote epitaxy by leveraging the conductivity and step-edge density of substrates with non-polar bonding. Various non-polar remote epitaxies are demonstrated, including Au film, $Cu_2O$ film and ZnO nanorods on ML-graphene/vicinal Au substrate, Cu film on ML-graphene/Si wafer, and Cu film on graphene-Cu foil. Experimental and computational studies indicate that remote interactions arise from electron depletion across the vdW interlayer, induced by coupling between graphene and the conductive vicinal steps. These remote epilayers possess excellent crystallinity and coherent epitaxial registry with their underlying substrates with non-polar bonding, as shown by X-ray scattering and electron microscopy. Using Cu foil as an example, we practically apply non-polar remote epitaxy on commercially available graphene-Cu foil to produce freestanding single-crystal Cu foils at room temperature and ambient pressure. Our work expands remote epitaxy from ionic to metallic and covalent-bonded materials and provides an innovative pathway in next-generation high-quality material manufacturing.

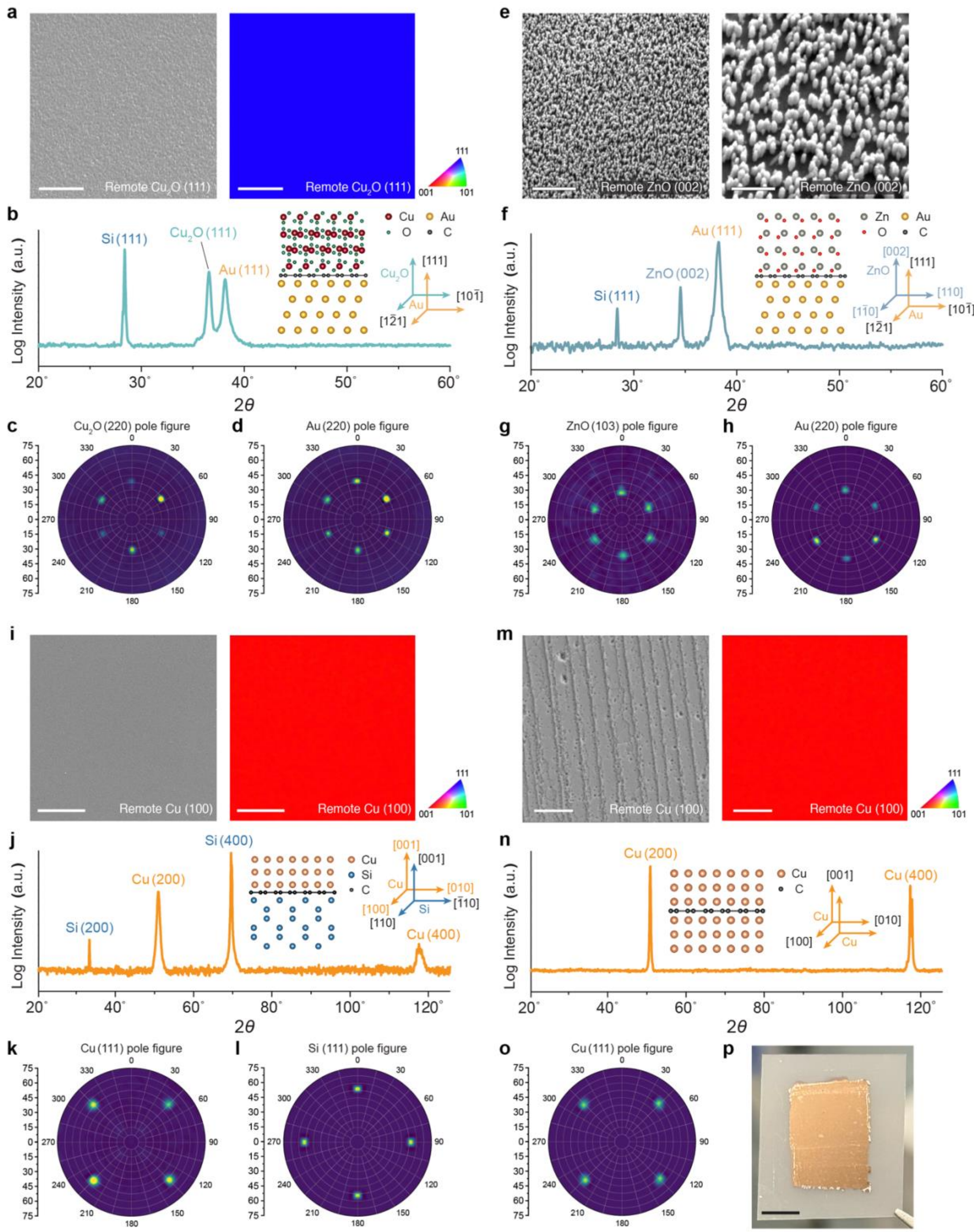

**Fig. 4. Demonstration of the universality of non-polar remote homo- and heteroepitaxy on covalent and metallic bonded substrates. a**, SEM image of $Cu_2O$/ML-graphene/vicinal-Au and the corresponding EBSD inverse pole figure mapping of the remote epitaxial $Cu_2O$ film. **b,** XRD

pattern (under the logarithmic scale) for $Cu_2O$ film on ML-graphene/vicinal-Au. **c**,**d**, X-ray pole figure from the $Cu_2O$ (220) (**c**) and Au (220) (**d**) facets. **e**, SEM images of ZnO nanorods on ML-graphene/vicinal-Au. **f**, XRD pattern (under the logarithmic scale) for ZnO nanorods on ML-graphene/vicinal-Au. **g**,**h**, X-ray pole figure from the ZnO (103) (**h**) and Au (220) (**i**) facets. **i**, SEM image of Cu film on ML-graphene/vicinal-As-doped Si and the corresponding EBSD inverse pole figure mapping of the remote epitaxial Cu film. **j,** XRD pattern (under the logarithmic scale) for Cu film on ML-graphene/vicinal-As-doped Si. **k**,**l**, X-ray pole figure from the Cu (111) (**k**) and Si (111) (**l**) facets. **m**, SEM image and the corresponding EBSD inverse pole figure mapping of the remote epitaxial of Cu film on graphene/Cu foil. **n**, XRD pattern (under the logarithmic scale) for Cu films on graphene/Cu foil. **o**, X-ray pole figure from the Cu (111) facets. **p**, Optical microscopic image of exfoliated Cu film with a size of 20 x 10 $mm^2$. Scale bars, 5 μm (**a**), 2 μm (**e**; left), 1 μm (**e**; right), 5 μm (**i**), 2 μm (**m**), 5 mm (**p**).

**Acknowledgments**

The authors gratefully acknowledge the staff at the APS beamline 9-ID – Dr. Hongrui He, Dr. Joseph Strzalka, Dr. Peco Myint, Dr. Miaoqi Chu, Dr. Zhang Jiang, Mr. Raymond Ziegler, and Dr. Suresh Narayanan for their support during preliminary GIWAXS experiments, which provided critical insights that shaped the direction of this research.

The project is sponsored by the startup fund of the Pritzker School of Molecular Engineering, University of Chicago. C.-T. F. acknowledges the financial support from Ministry of Education, Taiwan under the award number #1122501340B. P.-C. H. and P.-J. Hung gratefully acknowledge support from the National Science Foundation under award number NSF ECCS-2324286.

This research was performed on APS beam time award(s) (DOI: https://doi.org/10.46936/APS-188820/60013331, https://doi.org/10.46936/APS-189194/60013707, https://doi.org/10.46936/APS-189614/60013809, https://doi.org/10.46936/APS-190010/60014106, https://doi.org/10.46936/APS-190310/60014363, https://doi.org/10.46936/APS-190633/60014518, https://doi.org/10.46936/APS-190691/60014561, https://doi.org/10.46936/APS-191029/60014803, https://doi.org/10.46936/APS-191138/60014896, and https://doi.org/10.46936/APS-193288/60016373) from the Advanced Photon Source, a U.S. Department of Energy (DOE) Office of Science user facility operated for the DOE Office of Science by Argonne National Laboratory under Contract No. DE-AC02-06CH11357.

Work performed at the Center for Nanoscale Materials, a U.S. Department of Energy Office of Science User Facility, was supported by the U.S. DOE, Office of Basic Energy Sciences, under Contract No. DE-AC02-06CH11357.

This work made use of the Pritzker Nanofabrication Facility part of the Pritzker School of Molecular Engineering at the University of Chicago, which receives support from Soft and Hybrid Nanotechnology Experimental (SHyNE) Resource (NSF ECCS-1542205), a node of the National Science Foundation's National Nanotechnology Coordinated Infrastructure.

This work made use of the shared facilities at the University of Chicago Materials Research Science and Engineering Center, supported by the National Science Foundation under award number DMR-2011854.

This work made use of instruments in the Electron Microscopy Core of UIC's Research Resources Center. Acquisition and upgrade of the UIC JEOL JEM ARM200CF was supported by an MRI-R grant (DMR-0959470) and an MRI grant (DMR-1626065) from NSF.

## Author contributions

C.-T.F. and P.-C.H. conceived and developed the idea and planned the experiments. P.-J.H. and H.Z. designed and conducted synchrotron X-ray experiments and analyzed the data. X.L. and Y.S. performed DFT calculation and analysis. X.Z., I.K., and B.L. conducted synchrotron beamline configuration and assisted with the GIWAXS data acquisition. Q.F., Q.L. and X.W. assisted in GIWAXS measurements. S.M and J.R.G. performed the STM imaging. Y.H. C.X., T.H. and Z.Z. carried out EBSD experiments and VESTA crystallographic analysis. C.-T.F. performed the FIB. X.Z., F.S., and Y.L. performed the STEM imaging, ESD analysis and EELS data acquisition. A.S.F. assisted with the XRD data acquisition. C.-H.L. performed the SEM imaging. All authors analyzed the data and co-wrote the paper.

## Declaration of Interests

The authors declare no conflict of interest.

## Data and materials availability

All data are present in the manuscript and the Supplementary Materials. Additional data related to this paper are available from the corresponding authors upon reasonable request.